\documentclass[aps,pra,reprint,showpacs,superscriptaddress,twocolumn,10pt]{revtex4-1}
\usepackage{} %
\usepackage{amsfonts}
\usepackage{mathrsfs}
\usepackage{amsmath}
\usepackage{amssymb}
\usepackage{graphicx}
\usepackage{color}%
\usepackage{mathtools}
\usepackage{booktabs}
\usepackage{mleftright}

\usepackage{float}
\usepackage{graphicx}
\usepackage[colorlinks,linkcolor=blue,citecolor=blue,urlcolor=blue,hyperindex,bookmarks=false,pdfstartview=FitH]{hyperref}
\usepackage[utf8]{inputenc}

\setcitestyle{open={[},close={]},citesep={,\!},numbers}
\usepackage{color}

\begin{document}
\raggedbottom

\title{Quantum origin of diffraction from bright and dark states}

\author{Jian-Jian Cheng}
\affiliation{School of Science, Xi’an University of Posts and Telecommunications, Xi’an 710121, China}

\author{Jun-Ling Che}
\affiliation{School of Science, Xi’an University of Posts and Telecommunications, Xi’an 710121, China}

\author{Lin Zhang}
\email{zhanglincn@snnu.edu.cn}
\affiliation{School of Physics and Information Technology, Shaanxi Normal University, Xi'an 710119, China}

\author{Ming-Liang Hu}
\email{mingliang0301@163.com}
\affiliation{School of Science, Xi’an University of Posts and Telecommunications, Xi’an 710121, China}

\begin{abstract}
Building upon the recently introduced particle interpretation of the double-slit experiment [Phys. Rev. Lett. 134, 133603 (2025)] which attributes interference phenomena to detector-coupled (bright) and detector-uncoupled (dark) states of light, we develop a continuous-mode extension of the bright- and dark-state framework.  This extension addresses a conceptual distinction between interference and diffraction, that is, the transition from a finite set of discrete paths to a continuum of modes. Through the construction of a complete detector-oriented basis for single-slit diffraction, we demonstrate that the observed diffraction pattern arises from projection of the photon state onto a single bright mode by identifying the detectable and undetectable modes, with photons detected at intensity minima having zero probability, as they reside in modes spanning an infinite-dimensional dark subspace. Our approach thus provides a unified particle-based explanation of diffraction that connects quantum and classical wave optics, and reveals distinctive quantum signatures in higher-order correlations.	
\end{abstract}


\maketitle

\section{Introduction} \label{sec:1}
The nature of light has captivated scientific inquiry for centuries, with its wave-like and particle-like behaviors standing as central pillars of modern optics \cite{MandelWolf, ScullyZubairy}. Early foundational work, including Young's double-slit experiment and Fresnel's diffraction theory, firmly established the wave theory of light \cite{Rubinowicz1957,Young2019,Epstein1924}. The Huygens--Fresnel principle, in particular, became a cornerstone of the wave optics, accurately predicting complex light distribution by modeling wavefronts as collections of coherent secondary sources \cite{Huygens1912}. Yet, with the advent of quantum theory and the explanation of phenomena such as the photoelectric effect, the particle nature of light resurfaced, leading to the enduring concept of wave-particle duality \cite{Arons1965,MandelWolf,ScullyZubairy,Grangier1986,Daniele2010}.

This duality remains deeply puzzling when considering experiments at the single-photon level. Even when the photons are emitted one by one \cite{Somaschi2016}, they gradually build up diffraction and interference patterns, as though each indivisible quantum entity exhibits nonlocal and wave-like behavior \cite{MandelWolf,ScullyZubairy,Loudon2000,Kocsis2011,Rueckner2013, Luo2024, Fedoseev2025,Gibney2025,Zavatta2004,Parigi2007}. This gives rise to a conceptual tension: if light consists of particles, a photon passing through an aperture ought to have a nonzero probability of reaching any point on a detection screen \cite{ScullyZubairy}. Classical wave optics, however, predicts regions of exactly zero intensity due to destructive interference \cite{Rubinowicz1957}. Where, then, do the photons go in these dark regions?

A recent advance by Villas-Boas \textit{et al.} offers a compelling resolution within a discrete-mode model of interference \cite{VillasBoas2025,VillasBoas202502}. By extending Glauber’s quantum theory of optical coherence \cite{Glauber1, Glauber2}, they reframe the interference in terms of bright and dark collective states \cite{Dicke, Alsing1992, Bjork2001}. In this picture, a photon in the dark state does not vanish. It simply resides in a mode that is decoupled from the detector. This elegantly reconciles the particle and wave descriptions, showing that destructive interference corresponds not to the absence of photons, but to their presence in states that remain undetectable to a localized sensor \cite{Dicke, Alsing1992}.

While this discrete-mode approach provides a clear and intuitive picture for double- and multi-slit interference, diffraction from a single-slit or a general aperture presents a more fundamental and ubiquitous challenge \cite{Hiekkamaki2022}. Such systems are inherently continuous in nature, requiring a theoretical framework beyond the discrete path idealization \cite{MandelWolf, Fedoseev2025, Dodonov2005, ScullyZubairy}. This raises a pivotal question: are interference and diffraction fundamentally distinct phenomena, or do they originate from a common quantum principle?

At the quantum level, both interference and diffraction originate from the coherent addition of probability amplitudes, while differing in how these amplitudes are structured. Interference, as seen in the double-slit experiment, results from the discrete superposition of a finite set of paths. Diffraction, on the other hand, stems from the integration over a continuous wavefront through an aperture, corresponding to an infinite number of possible paths. This conceptual difference is reflected mathematically: interference can be described with a finite set of optical modes, while diffraction requires a continuous-mode formulation, for which a complete detector-oriented basis has not been fully established.

In this study, we extend the framework of bright and dark states into the continuous-mode regime and develop a quantum description of the single-slit diffraction. We will show that diffraction can be characterized by an infinite-dimensional dark subspace, which mathematically mirrors the continuous structure of the aperture. This framework provides an explanation for the absence of detections in dark regions by showing that photons occupy dark states are uncoupled to the detector. This yields a description of diffraction that bridges classical wave optics and quantum theory from a purely particle-based perspective.

\section{Glauber's framework for quantum diffraction} \label{sec:2}
In accordance with Glauber’s theory of quantum optical coherence, we model the measurement process at a microscopic level through energy exchange between the radiation field and a sensor atom. Within this framework, the quantum statistical properties of the field are captured by the electrical field operator \cite{Glauber1, Glauber2}:
\begin{equation}
 \hat{E}(\mathbf{r},t) = \hat{E}^{(+)}(\mathbf{r},t) + \hat{E}^{(-)}(\mathbf{r},t),
\end{equation}
where $\hat{E}^{(+)}(\mathbf{r},t)$ and $\hat{E}^{(-)}(\mathbf{r},t)$ represent the positive and negative frequency components, respectively, and quantum mechanically, they correspond to the photon annihilation and creation operators, i.e., $\hat{E}^{(+)}(\mathbf{r},t) \propto \hat{a}$ and $\hat{E}^{(-)}(\mathbf{r},t) \propto \hat{a}^\dagger$.

The probability of detecting a photon from a single-mode field in the state $|\Psi\rangle$ is proportional to the expectation value
\begin{equation}
 \langle\Psi| \hat{E}^{(-)}(\mathbf{r},t) \hat{E}^{(+)}(\mathbf{r},t) |\Psi\rangle \propto \langle\Psi| \hat{a}^\dagger \hat{a}|\Psi\rangle,
\end{equation}
which arises naturally from the energy-exchange mechanism between the field and the sensor atom. This interaction is described under the rotating-wave approximation by the Hamiltonian \cite{ScullyZubairy}
\begin{equation} \label{Hami}
 \hat{H} = \hat{E}^{(+)}(\mathbf{r},t) \hat{\sigma}^{+} + \hat{E}^{(-)}(\mathbf{r},t) \hat{\sigma}^{-},
\end{equation}
where $\hat{\sigma}^{+}$ ($\hat{\sigma}^{-}$) is the raising (lowering) operator that induces transition between the ground state $|g\rangle$ and excited state $|e\rangle$ of the sensor atom.

In contrast to the classical treatment of the single-slit diffraction via the Huygens--Fresnel principle which constructs the wavefront through superposition of the elementary spherical subwaves, we develop a quantum formulation based on Glauber's theory of optical coherence. Here, each $x \in [0, b]$ corresponds to a spatial mode operator $\hat{a}(x)$ satisfying $[\hat{a}(x), \hat{a}^\dagger(x')] = \delta(x-x')$, and the single-photon state at the position $x$ is $|1_x\rangle = \hat{a}^\dagger(x)|0\rangle$. These states constitute a complete basis, i.e., $\langle 1_x | 1_{x'}\rangle = \delta(x-x')$ and $\int_0^b |1_x\rangle\langle 1_x| dx = \hat{I}$, with $\hat{I}$ being the identity operator. In quantum optics, the field operator can be expanded in a complete set of orthogonal field modes. Here, the coordinate $x$ serves as a continuous classical variable labeling a specific set of the spatial modes across the single slit. This approach follows Glauber’s theory \cite{Glauber1, Glauber2} and connects naturally to the Huygens--Fresnel picture which takes the wavefront as a collection of the coherent secondary sources. Each such source corresponds to an independent spatial mode, for which $\hat{a}^\dag(x)$ [$\hat{a}(x)$] creates (annihilates) a photon in the mode labeled by $x$, and $|1_x\rangle$ represents a photon occupying that mode. Moreover, $\{|1_x\rangle\}_{x\in[0,b]}$ is the set of basis states in this mode space and $\langle 1_{x} | 1_{x'} \rangle = \delta(x - x')$ indicates that a photon definitely in mode $x$ has zero probability to be in mode $x'$. Whereas the classical Huygens--Fresnel principle sums the field amplitudes, our quantum construction builds the optical state from a coherent superposition of $|1_x\rangle$, thereby establishing a fundamentally particle-based description of diffraction \cite{Glauber1, Glauber2}.

\begin{figure}[htbp]
\centering
\includegraphics[width=230pt]{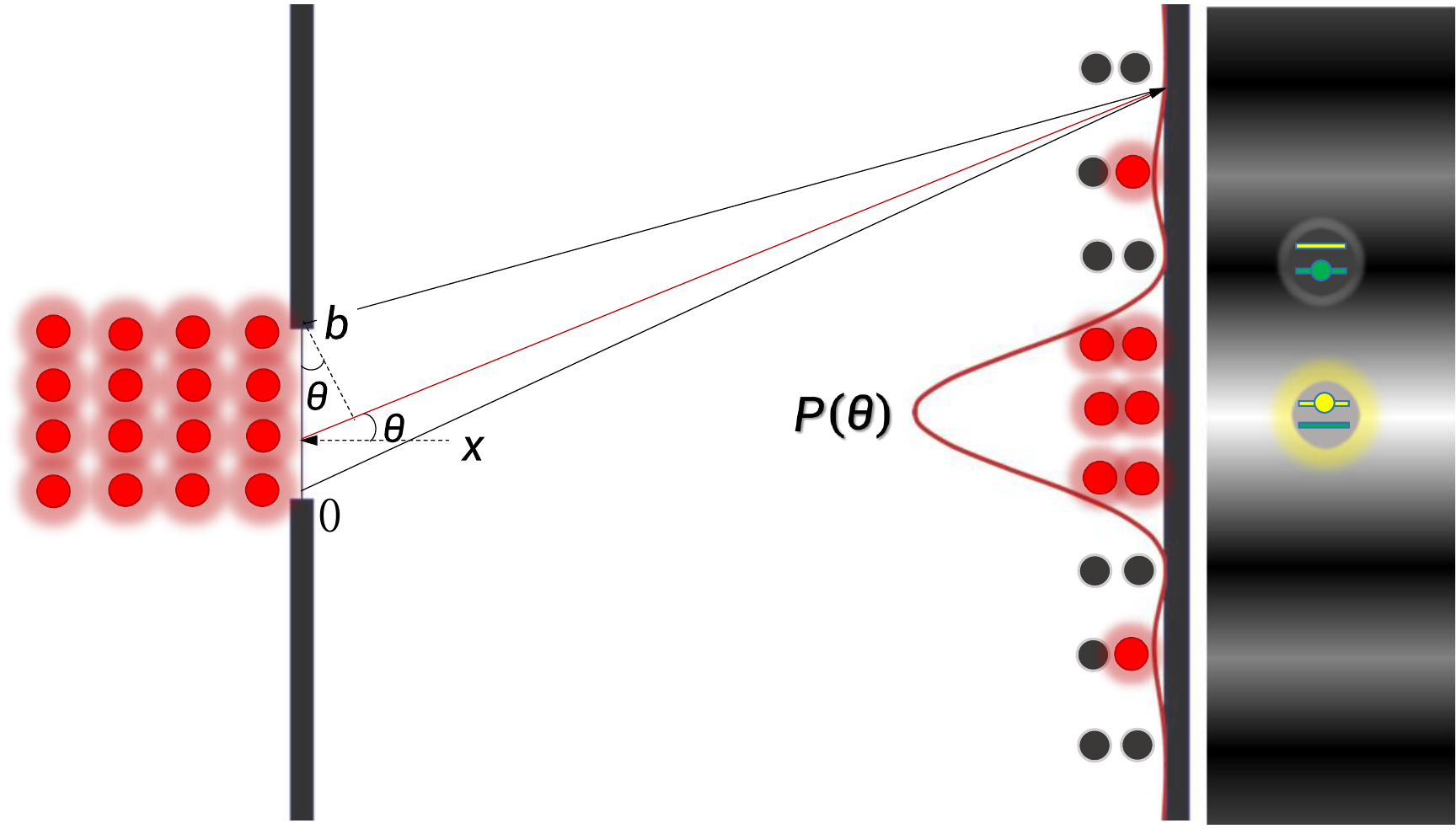}
\caption{Fraunhofer single-slit diffraction intensity distribution, where the red envelope represents the diffraction pattern predicted by classical wave optics. The quantum interpretation is visualized as follows: the red dots on the screen represent photons in bright states, which can excite the sensor atoms and thereby cause a detectable response from the detector. In contrast, the gray dots correspond to photons in dark states which fail to excite the sensor atoms; hence although physically present, these photons remain undetectable.} \label{Fig}
\end{figure}

We accordingly construct the positive-frequency part of the field operator for direction $\theta$ as (see Fig. \ref{Fig})
\begin{equation} \label{eq4}
 \hat{\mathcal{E}}^{(+)}(\theta) = \frac{1}{\sqrt{b}} \int_0^b \hat{a}(x) e^{i\phi(x,\theta)} dx.
\end{equation}
This defines a collective mode, described by the operator $\hat{\mathcal{E}}^{(+)}(\theta)$ which obeys $[\hat{\mathcal{E}}^{(+)}(\theta), \hat{\mathcal{E}}^{(-)}(\theta)] = 1$. This commutation relation underpins the detector-oriented basis. In the far-field limit, the spherical wavefronts emitted from different points of the slit can be approximated as plane waves, with phases that depend linearly on the position. So $\phi(x,\theta) = \frac{2\pi}{\lambda} x \sin\theta$ represents the phase delay under the Fraunhofer approximation for a wave originating at point $x$ of the slit and detected in the $\theta$ direction.

The interaction between the radiation field and a localized detector, a sensor modeled as a two-level atom, is governed by the light-matter interaction Hamiltonian
\begin{equation}
 \hat{H} = g \hat{\mathcal{E}}^{(+)}(\theta)  \sigma^{+} + \text{H.c.},
\end{equation}
where $g=g(\mathbf{r})$ denotes the position-dependent coupling strength, taken as uniform across the relevant modes for simplicity, and $\mathbf{r}$ specifies the location of the sensor atom. This Hamiltonian embodies a foundational principle of quantum measurement: the detector does not respond to the total photonic field, but only to that particular collective mode defined by the positive-frequency operator $\hat{\mathcal{E}}^{(+)}(\theta)$. A photon in a field state can induce a transition from the atomic ground state $|g\rangle$ to the excited state $|e\rangle$ only if the photon state possesses a nonzero component along this specific mode. Conversely, any field component that lies in the kernel of $\hat{\mathcal{E}}^{(+)}(\theta)$, forming a dark state, remains completely decoupled from the sensor atom. This formulation offers an interpretation of directional diffraction within a quantum-mechanical framework, where the detection probability is governed by the intrinsic properties of the sensor atom and the photon state.

\section{Detector-oriented basis for single-photon diffraction} \label{sec:3}
In the Fraunhofer diffraction regime with normal incidence, the incident plane wave illuminates the slit uniformly, resulting in a symmetric photonic state:
\begin{equation} \label{eq6}
  |S\rangle =\frac{1}{\sqrt{b}} \int_0^b |1_x\rangle dx,
\end{equation}
where $|1_x\rangle$ represents the basis state of a single photon associated with $x$, and $|S\rangle$ describes the photon’s distribution under uniform illumination. In the classical picture, uniform illumination is described by a constant field amplitude; in the quantum picture, by contrast, it corresponds to an equal-probability coherent superposition of $|1_x\rangle$ across $x\in[0,b]$. This superposition forms a delocalized single-photon state whose spatial profile reproduces the classical diffraction pattern in the far field.

Glauber's theory of optical coherence successfully predicts detection probabilities through the first-order correlation function. Specifically, the probability for a detector to absorb a single photon at an angle $\theta$ is governed by \cite{Glauber2}
\begin{equation}
 G^{(1)}(\theta) = \langle \psi | \hat{E}^{(-)} \hat{E}^{(+)} | \psi \rangle
                 = | \langle 0 | \hat{E}^{(+)} | \psi \rangle |^2,
\end{equation}
which quantifies the likelihood of exciting a sensor atom placed at that angular position. While this approach correctly predicts the far-field diffraction pattern, it does not explicitly reveal the underlying quantum mechanism that determines which components of the optical state couple to the detector. In particular, for single-slit diffraction, it remains unclear how a localized sensor atom selectively interacts with a specific collective mode among the continuum emerging from the single slit, and how the spatial relationship between the detector and the photon state governs the energy exchange.

To address the aforementioned questions, we develop a quantum description of single-slit diffraction by introducing a detector-oriented basis. The central idea is to construct a complete orthonormal set of states $\{|\psi_n(\theta) \rangle \}_{n \in \mathbb{Z}}$ associated with the detection direction $\theta$ (see Fig. \ref{Fig}). These detector-oriented basis states are defined as
\begin{equation} \label{eq8}
 |\psi_n(\theta)\rangle = \frac{1}{\sqrt{b}} \int_{0}^{b} e^{-i\phi(x,\theta)} e^{i k_n x} |1_x\rangle dx,
\end{equation}
with $k_n = {2\pi n}/{b}$. Here, the phase factor $e^{-i\phi(x,\theta)}$ compensates for the propagation phase from the slit to the detector, thereby linking the basis intrinsically to the detector's directional sensitivity, while $e^{i k_n x}$ provides a Fourier modulation that ensures orthogonality. As shown in Appendix \ref{sec:A}, $\{|\psi_n(\theta)\rangle\}$ constitute a complete basis, so any single-photon state can be spanned in this basis.

The $n = 0$ state is the bright state $|\text{Bright}_\theta\rangle = |\psi_0(\theta)\rangle$, which satisfies (see Appendix \ref{sec:B})
\begin{equation} \label{eq9}
 \hat{\mathcal{E}}^{(+)}(\theta) |\text{Bright}_\theta\rangle = |0\rangle,
\end{equation}
which reflects a match between the state and a detectable response from a sensor. As a result, a photon in $|\text{Bright}_\theta\rangle$ has the highest probability of being absorbed by a sensor atom in the $\theta$ direction, corresponding to the strongest light-matter coupling.

For $n \neq 0$, the basis states $|\psi_n(\theta)\rangle$ correspond to the dark modes $|\text{Dark}_{n,\theta}\rangle$ that satisfy (see Appendix \ref{sec:B})
\begin{equation} \label{eq10}
 \hat{\mathcal{E}}^{(+)}(\theta) |\text{Dark}_{n,\theta}\rangle = 0.
\end{equation}
Although the photons can occupy these states, their specific phase structure associated with $k_n$ makes them remain decoupled from the detector under the interaction Hamiltonian $\hat{H}$. As a result, they are unable to excite the sensor atom in the $\theta$ direction.

Since the states $\{|\psi_n(\theta)\rangle\}$ constitute a complete basis, for which $|\psi_0(\theta)\rangle$ and \(\{|\psi_n(\theta)\rangle\}_{n\neq 0}\) are physically identified as the bright and dark states, respectively, the single-photon state $|S\rangle$ in Eq. \eqref{eq6} can be naturally spanned as
\begin{equation} \label{eq14}
 |S\rangle= c_0 |\text{Bright}_\theta\rangle + \sum_{n \neq 0} c_n |\text{Dark}_{n,\theta}\rangle,
\end{equation}
where $c_n=\langle \psi_n(\theta)|S\rangle$ ($\forall n$), and from Eqs. \eqref{eq6} and \eqref{eq14} one can obtain that $c_n = [e^{i(\beta - n\pi)} \sin(\beta - n\pi)]/(\beta - n\pi)$, where $\beta = (\pi b \sin\theta)/\lambda$ (see Appendix \ref{sec:C}). Here, the normalization condition $\sum_{n} |c_n|^2 = 1$ ensures that the probability is conserved, confirming that the photon states are entirely redistributed among the basis states without loss. This expansion recasts the diffraction process in terms of projection probabilities onto detector-defined bright and dark states. The correspondence with classical wave optics becomes evident in the bright-state projection. The detection probability $|c_0|^2 = [\sin(\beta)/\beta]^2$ exactly reproduces the classical single-slit diffraction profile $P(\theta)$. While the classical theory attributes this pattern to the interference of field amplitudes, the quantum description identifies it with the probability of detecting a photon that is in the bright mode, as illustrated in Fig.~\ref{Fig}. Moreover, our continuum model reveals an infinite-dimensional dark subspace, a direct consequence of the slit's finite width. This richness in dark states originates from the Fourier link between position at the slit and direction in the far field: any wave vector orthogonal to the bright mode forms a distinct dark state. Thus, a diffraction minimum simply indicates that if a photon is present, it would be found in the dark-state subspace, making it undetectable by a sensor coupled only to the bright mode. This unitary and basis-dependent description conserves photon number perfectly, demonstrating that at the dark regions, photons are not absent but rather undetectable, and that the diffraction pattern itself is an imprint of the detection process.

In a single-slit diffraction system, the total photon number operator is defined by integrating over the entire slit width: $\hat{N}_{\text{total}} = \int_0^b \hat{a}^\dagger(x) \hat{a}(x)  dx$. This operator represents a global property of the system, yielding the average number of photons in the entire quantum state. For a uniformly illuminated single-photon state $|S\rangle$, its expectation value $\langle S | \hat{N}_{\text{total}} | S \rangle = 1$, ensuring that one and only one photon is present in the system. Although the detection probability varies with angle to form the diffraction pattern, the total number of photons in the system is strictly conserved. The actual detection process, being local in nature, is governed by direction-dependent field operator expectation values, such as the first-order correlation function $G^{(1)}(\theta)$. Hence, while $\langle \hat{N}_{\text{total}} \rangle = 1$ describes the statement that “there is one photon in the system”, $G^{(1)}(\theta)$ answers the question: “what is the probability of detecting that photon at angle $\theta$?” (see Sec. \ref{sec:4} for more details) Even when switching from the position basis to the detector-oriented bright-dark state basis, this global conserved quantity remains unity. At the diffraction minima where the detection probability drops to zero, any photon present would be found in the dark states that are decoupled from the detector rather than being absent from the system.

Furthermore, we examine the structure of the dark-state subspace. The weights of the basis states, $|c_n|^2$, decay rapidly with increasing mode index $|n|$, mirroring the suppression of high spatial frequencies in classical diffraction. For example, at $\beta = 0.3\pi$, the bright state accounts for $|c_0|^2 \approx 0.737$ of the total probability, leaving only 26.3\% distributed among dark states. A closer inspection reveals a strongly inhomogeneous internal structure: the $n = 1$ mode dominates, contributing about 51.4\% of the dark components ($|c_1|^2 \approx 0.135$), followed by the $n = -1$ mode at roughly 14.9\% ($|c_{-1}|^2 \approx 0.039$). The $n = 2$ and $n = -2$ modes contribute approximately 8.7\% and 4.8\%, respectively. In contrast to the double-slit case, for which only one dark state exists, this hierarchical distribution demonstrates that the dark-state subspace in single-slit diffraction possesses a well-defined internal stratification, dominated by low-wavenumber modes.

The rapid decay of $|c_n|^2$ with the increasing $|n|$ confirms that, although the dark-state subspace is infinite-dimensional, the wave function is effectively concentrated in the bright mode and a few low-order dark modes. This structure aligns with the decay behavior of the Fourier components in classical diffraction, highlighting a profound correspondence between the construction of quantum states and classical wave optics. The internal stratification of the dark subspace further clarifies that the non-detectability at the diffraction minima stems from the orthogonality between the photon and sensor states \cite{Alsing1992, Dodonov2005}, rather than from the physical absence of the photon. This represents a fundamental transition from modeling double-slit interference with discrete modes to describing single-slit diffraction within a continuous-mode framework. While both are unified within the picture of the bright and dark states, they exhibit distinctly different internal structures due to the fundamental difference in the dimensionality of their state spaces.

\section{Nonclassical Signatures in Multi-Photon Diffraction} \label{sec:4}
As discussed above, dark states correspond to eigenstates of the operator $\hat{\mathcal{E}}^{(+)}(\theta)$ with zero eigenvalue, which suggests that they may remain undetectable by conventional sensors such as two-level atoms.   Then a natural question that arises is to what extent can quantum mechanical bright and dark states account for the classical phenomenon of diffraction? We now extend the detector-dependent mode eigenstates to an operator formalism that accommodates arbitrary photon numbers. The creation operators are defined through the fundamental relation
$\hat{J}_n^\dagger |0\rangle = |\psi_n(\theta)\rangle$,
which leads directly to
\begin{equation} \label{Jn}
 \hat{J}_n^\dagger =\frac{1}{\sqrt{b}} \int_{0}^{b}  e^{-i\phi(x,\theta)} e^{i k_n x} \hat{a}^\dagger(x) dx,
\end{equation}
and it is direct to show that these operators obey bosonic commutation relation $[\hat{J}_m, \hat{J}_n^\dagger] = \delta_{mn}$ (see Appendix \ref{sec:D}). This ensures that each $\hat{J}_n^\dagger$ creates the single-photon basis state $|\psi_n(\theta)\rangle$ from the vacuum state, thereby shifting the continuous spatial mode operators $\{\hat{a}^\dagger(x)\}$ into the orthogonal mode operators $\{\hat{J}_n^\dagger\}$ that are adapted to the detection direction $\theta$. The construction of these orthogonal modes is analogous to that of the field-orthogonal temporal modes of photonic states in the temporal-mode framework, in which the temporal-mode operators obey $[\hat{A}_i, \hat{A}_j^\dagger] = \delta_{ij}$ \cite{Brecht2015}. In this sense, our framework is reminiscent of the temporal-mode framework, but they address different physical problems: the temporal modes have a distinct advantage for quantum information processing, and the orthogonal modes in our framework are introduced for explaining the diffraction phenomenon.

The total photon number operator $\hat{N}_{\text{total}}$ then satisfies $\hat{N}_{\text{total}} = \sum_n \hat{J}_n^\dagger\hat{J}_n$, manifesting the conservation of photon number in diffraction. For $n = 0$, $\hat{B}_\theta^\dagger= \hat{J}_0^\dagger$ defines the bright-state operator, which creates photon in state that excites the sensor atom in the $\theta$ direction and thus can be detected. For $n \neq 0$, the dark-state operators $\hat{D}_{n,\theta}^\dagger=\hat{J}_n^\dagger$ create photons in states that are unable to excite the sensor atom and thus cannot be detected. The extension to multi-photon states provides a robust demonstration of how the bright- and dark-mode formalism accounts for the classical diffraction phenomenon \cite{VillasBoas2025, ScullyZubairy}. This framework offers a unified description bridging quantum and classical regimes, revealing how diffraction patterns emerge from the statistical distribution of photons across the detector-defined modes.

In classical wave optics, the Huygens--Fresnel principle successfully predicts the intensity distribution of diffraction patterns by treating each point on a wavefront as a source of secondary waves. Their interference produces the alternating bright and dark fringes, answering the fundamental question of where light intensifies and where it diminishes. Within the above quantum framework, the particle-based explanation of optical diffraction has largely relied on two equivalent descriptions. One derives from the ensemble average of many single-photon detection events, while the other comes from the statistical interpretation of a single measurement performed on a multi-photon state. Although both theories predict identical diffraction patterns, they jointly obscure an important theoretical subtlety: how different multi-photon quantum states, such as Fock states and coherent states, each produces the same observed diffraction phenomenon through their distinct quantum statistical properties.

To properly describe the multi\-photon state in this continuous setting, we first introduce the global creation operator $\hat{A}^{\dagger} \equiv \frac{1} {\sqrt{b}} \int_0^{b} \hat{a}^{\dagger}(x) \, dx$, which obeys the bosonic commutation relation $[\hat{A}, \hat{A}^{\dagger}] = 1$. This definition stems from the physical picture of uniform illumination across the slit in classical Fraunhofer diffraction: $\hat{A}^\dagger$ corresponds to a collective mode whose spatial profile is uniform over the entire aperture, i.e., $\hat{A}^\dagger|0\rangle = |S\rangle$. The term “global” thus emphasizes that this mode spans the whole slit with a homogeneous profile. At the same time, this construction offers a convenient formulation for multi-photon states, and the $N$-photon Fock state under uniform illumination can be written compactly as
\begin{equation}
 |N\rangle = \frac{1}{\sqrt{N!}} (\hat{A}^\dagger)^N |0\rangle.
\end{equation}
A central feature of this construction is that $\hat{A}^\dagger$ can be decomposed as (see Appendix \ref{sec:E})
\begin{equation} \label{eq17}
 \hat{A}^\dagger = c_0 \hat{B}_\theta^\dagger + \sum_{n \neq 0} c_n \hat{D}_{n,\theta}^\dagger,
\end{equation}
and this expansion shows that the creation of a uniformly distributed photon via $\hat{A}^\dagger$ coherently distributes the photon among the bright mode and a set of dark modes, with the weights $|c_n|^2$ dictated by the detection angle $\theta$. Here, the operators $\hat{B}_\theta^\dagger$ and $\hat{D}_{n,\theta}^\dagger$ extend the detector-oriented basis $\{|\psi_n(\theta)\rangle\}$ originally defined for a single photon into the multi-photon regime. It thereby inherits the essential physical property that only the bright mode couples to a sensor atom in the $\theta$ direction, while all the dark modes remain completely decoupled.

To analyze the detection process, we consider the action of the bright-state operator $\hat{B}_\theta$ on $|N\rangle$. Using the orthogonality $[\hat{B}_\theta, \hat{D}_{n,\theta}^\dagger] = 0$ and the relation $[\hat{B}_\theta, \hat{A}^\dagger] = c_0$, we obtain the generalized commutator
\begin{equation}
 [\hat{B}_\theta, (\hat{A}^\dagger)^N] = N c_0 (\hat{A}^\dagger)^{N-1}.
\end{equation}
Applying this to the vacuum and using the normalization condition of Fock states leads to
\begin{equation}
 \hat{B}_\theta |N\rangle = \sqrt{N} c_0 |N-1\rangle,
\end{equation}
which describes the annihilation of one photon from the $N$-photon state, with amplitude $\sqrt{N} c_0$. Within the light-matter interaction model, this result implies
\begin{equation}
 \hat{H} |N\rangle |g\rangle = g \sqrt{N} c_0 |N-1\rangle |e\rangle.
\end{equation}

This expression captures the joint process of atomic excitation and single-photon absorption, with a transition amplitude controlled by $\sqrt{N} c_0$. The $\sqrt{N}$ factor reflects bosonic enhancement due to photon indistinguishability, while $c_0$ embodies the directional selectivity of the bright mode \cite{Dicke, MandelWolf}. These dynamics are analogous to those of the $N$-excitation Dicke model \cite{Dicke}, where collective states exhibit enhanced emission.

The mode occupation statistics further illustrates the above picture. In the detector-oriented basis, the expectation values take the form
\begin{align}
  \langle N| \hat{B}_\theta^\dagger \hat{B}_\theta |N\rangle  = N |c_0|^2, \quad
  \langle N| \hat{D}_{n,\theta}^\dagger \hat{D}_{n,\theta} |N\rangle  = N |c_n|^2.
\end{align}
This distribution reflects the quantum behavior of indistinguishable bosons: each photon is found in the bright mode with probability $|c_0|^2$, leading to an average bright-mode population of $N|c_0|^2$.  The photon state maintains a definite total photon number $\langle N| \hat{N}_{\text{total}}|N\rangle = N$, which strictly conserves the photon number in the system during diffraction. Accordingly, the first-order correlation function becomes
\begin{equation}
\begin{aligned}
 G^{(1)}_{\text{Fock}}(\theta) &= g^2 \langle N|\hat{\mathcal{E}}^{(-)}(\theta) \hat{\mathcal{E}}^{(+)}(\theta)|N\rangle \\
	                           & = g^2 N \left( \frac{\sin\beta}{\beta} \right)^2.
\end{aligned}
\end{equation}
Thus, for the Fock state, the classical diffraction profile, scaled by $N$, stems directly from the quantum probability amplitude associated with the bright state. While the above first-order correlation function captures the diffraction intensity pattern, the second-order correlation function may expose the underlying quantum statistics of the source. This difference reflects a deeper physical implication of the field operator $\hat{\mathcal{E}}^{(+)}(\theta)$: it represents not just a local field amplitude, but the entire far field mode detected at angle $\theta$. While the decomposition into bright and dark modes shapes the diffraction profile, the higher-order correlations reveal the intrinsic quantum nature of the optical state.

We now evaluate the second-order correlation function for the $N$-photon Fock state:
\begin{align}
 G^{(2)}_{\text{Fock}}(\theta) &= \frac{\langle N|\hat{\mathcal{E}}^{(-)}(\theta) \hat{\mathcal{E}}^{(-)}(\theta) \hat{\mathcal{E}}^{(+)}(\theta) \hat{\mathcal{E}}^{(+)}(\theta) |N\rangle}
                                  {\langle N| \hat{\mathcal{E}}^{(-)}(\theta) \hat{\mathcal{E}}^{(+)}(\theta) |N\rangle^2} \nonumber \\
	                           &= 1 - \frac{1}{N}.
\end{align}
This result unambiguously reveals photon antibunching, a hallmark of nonclassical light \cite{Grangier1986, Zavatta2004}. When $N=1$, the correlation function drops to zero, corresponding to perfect antibunching, whereas for large $N$ it approaches the coherent-state limit. The fact that such a nonclassical behavior emerges naturally within the formalism of bright and dark states highlights its broader value: it not only accounts for classical diffraction profiles but also offers a unified framework for analyzing how quantum statistics manifests in the wave-optical phenomena. These insights suggest a promising direction for future study of quantum diffraction in more complex optical systems, where the coupling between structured modes and quantum correlations could give rise to other nonclassical effects.

\section{Coherent States as the Quantum Origin of Classical Diffraction} \label{sec:5}
For a coherent state $|\alpha\rangle$ with uniform amplitudes (i.e., $\alpha(x) \equiv \alpha$), we begin by applying the displacement operator $\hat{D}(\alpha)$ to the vacuum, and this yields $|\alpha\rangle = \hat{D}(\alpha) |0\rangle$, with
\begin{equation} \label{eq24}
 \hat{D}(\alpha)= \exp\left(\int_0^b \left[\alpha \hat{a}^\dagger(x) - \alpha^* \hat{a}(x)\right] dx \right),
\end{equation}
and for analyzing the coupling of the photons in the state $|\alpha\rangle$ with the sensor atoms, we decompose $\hat{D}(\alpha)$ using $\hat{J}_n^\dagger$ defined in Eq. (\ref{Jn}), which yields (see Appendix \ref{sec:E})
\begin{equation} \label{eq25}
 \hat{D}(\alpha) = \exp\left( \sum_n \left[ \alpha c_n \sqrt{b} \hat{J}_n^\dagger - \alpha^* c_n^* \sqrt{b} \hat{J}_n \right] \right),
\end{equation}
and this transforms $\hat{D}(\alpha)$ from the position space into the space spanned by the detector-oriented basis $\{|\psi_n(\theta)\rangle\}$. Thereby, the integral over $x$ is explicitly decomposed into a sum over the bright- and dark-state operators.

Then a useful simplification becomes possible due to the commutation relation $[\hat{J}_m^\dagger, \hat{J}_n] = 0$ for $m \neq n$, which confirms the independence of different modes. This allows us to factorize the exponential into a product over modes:
\begin{equation}
 |\alpha\rangle = \prod_n \exp\left( \alpha c_n \sqrt{b} \hat{J}_n^\dagger - \alpha^* c_n^* \sqrt{b} \hat{J}_n \right) |0\rangle,
\end{equation}
given that the vacuum state $|0\rangle = \otimes_n |0\rangle_n$ and that each exponential acts only within its corresponding mode subspace, $|\alpha\rangle$ simplifies further to the fully factorized form
\begin{align}
  |\alpha\rangle &= \otimes_n \exp\left( \alpha c_n \sqrt{b} \hat{J}_n^\dagger - \alpha^* c_n^* \sqrt{b} \hat{J}_n \right) |0\rangle_n \nonumber \\
	             &= \otimes_n |\alpha_n\rangle.
\end{align}
where the direct product $\otimes_n$ signifies that the total state decomposes into a tensor product of the independent $|\alpha_n\rangle$ for each mode $n$. Each $|\alpha_n\rangle$ is a coherent state with amplitude $\alpha_n = \alpha c_n \sqrt{b}$, satisfying $\hat{J}_n |\alpha_n\rangle = \alpha_n |\alpha_n\rangle$. This factorization underscores a key property of the coherent states, i.e., the photons in different modes remain entirely uncorrelated \cite{Glauber1, MandelWolf}.

We can therefore write the total state explicitly as
\begin{equation} \label{eq25}
 |\alpha\rangle = |\alpha_{0}\rangle \otimes_{n\neq 0} |\alpha_{n}\rangle,
\end{equation}
which distinguishes the bright and dark modes. It follows that the bright- and dark-mode operators act as (see Appendix \ref{sec:F})
\begin{equation} \label{eq26}
  \hat{B}_\theta |\alpha\rangle = \alpha_0 |\alpha\rangle, \quad
  \hat{D}_{n,\theta} |\alpha\rangle = \alpha_n |\alpha\rangle,
\end{equation}
showing that only photons in the bright mode are coupled to the detector. Then the first-order correlation function reduces to
\begin{align} \label{eq27}
 G^{(1)}_{\text{Coh.}}(\theta) &= g^2 \langle\alpha|\hat{\mathcal{E}}^{(-)}(\theta) \hat{\mathcal{E}}^{(+)}(\theta)|\alpha \rangle \nonumber \\
	                           &= g^2 |\alpha|^2 b  \left( \frac{\sin \beta}{\beta} \right)^2,
\end{align}
with $|\alpha|^2 b$ being the total incident photon flux. The total average photon number, given by $\langle \alpha | \hat{N}_{\text{total}} | \alpha \rangle = |\alpha|^2 b$, is conserved and distributed between the bright and dark modes. This reproduces the classical diffraction pattern, confirming that the detected signal originates solely from photons in the bright mode, even though photons are also present in dark modes.

The bright- and dark-mode decomposition further exposes a fundamental difference between Fock and coherent states that is not visible in the first-order measurements. While these two kinds of states produce the same diffraction pattern, the underlying quantum structures are markedly different. The coherent states factorize independently across modes, while the Fock states exhibit strong intermode correlations due to the fixed total photon number. This structural distinction is clearly reflected in the second-order correlation function:
\begin{equation}
  G^{(2)}_{\text{Coh.}}(\theta)  = \frac{\langle \alpha| \hat{\mathcal{E}}^{(-)}(\theta) \hat{\mathcal{E}}^{(-)}(\theta) \hat{\mathcal{E}}^{(+)}(\theta) \hat{\mathcal{E}}^{(+)}(\theta)|\alpha \rangle}
                                   {\langle \alpha| \hat{\mathcal{E}}^{(-)}(\theta) \hat{\mathcal{E}}^{(+)}(\theta) |\alpha\rangle^2}
	                             = 1,
\end{equation}
which is characteristic of the Poissonian statistics and the absence of photon correlations in a coherent state.

The correspondence with the classical wave-optical result thus originates from the intrinsic nature of the coherent states \cite{Glauber1,ScullyZubairy,Steuernagel2001}. The Poissonian statistics and mode independence work together to suppress nonclassical features, allowing the quantum description to converge to the classical limit. Importantly, this classical-like behavior is inherent to coherent states and does not rely on the external decoherence.

\section{Outlook and conclusion} \label{sec:6}
In this work, we seek to contribute a particle-based description of diffraction. Grounded in a continuous-mode framework of bright and dark states, our approach offers a perspective on foundational questions and may help to motivate future experimental studies and potential applications \cite{VillasBoas2025,Blais2021,Phay2008,Murr2006,Steuernagel2001}. On the experimental front, platforms such as superconducting circuits and trapped-ion arrays offer a controlled setting to emulate a discretized version of the continuous slit using coupled electromagnetic or vibrational modes \cite{Blais2021}. Quantum state tomography of the multimode field could then directly reveal the mode occupation distribution: at diffraction minima, the bright-mode population vanishes while the total excitation remains unity, fully distributed among those dark modes, providing clear evidence for the dark space. Beyond foundational verification, the dark space offers intriguing practical prospects. As an infinite-dimensional subspace, it extends beyond the single dark state of discrete models, providing a naturally decoherence-free photonic memory. Encoding quantum information in this subspace protects it from radiative loss into the bright channel. More profoundly, the high dimensionality of the dark space allows redundant encoding across multiple dark modes, opening a path toward fault-tolerant quantum memory in structured photonic environments \cite{Cheng2021,Hu2018,Brekenfeld2020, C. J. Villas-Boas2024,Cheng2025}.

In summary, we have constructed a continuous-mode framework for single-slit diffraction. It extends Glauber's coherence theory into a particle-based interpretation of the diffraction pattern. By introducing a complete basis to characterize the detector's directional response, we established a rigorous framework, within which we showed that at every detection angle $\theta$, the optical field could be decomposed into a single bright mode and an infinite-dimensional dark mode. This decomposition reveals that the diffraction profile corresponds to population of photons in the bright mode, thereby linking wave-optical intensity patterns to the probability of detector excitation. Crucially, at the dark regions, the photons are not absent but rather occupy the structured dark modes, remaining physically undetectable as the sensor atom cannot be excited in this situation. We also demonstrated that while the Fock states exhibit spatially uniform antibunching across the diffraction pattern, the coherent states factorize completely into independent bright and dark modes. This highlights how the different quantum statistics manifest themselves in the same diffraction profile. All in all, by developing a continuous-mode framework of bright and dark states, we have provided a consistent and intuitive quantum explanation of diffraction, fully bridging the particle behaviors of photons and the wave-like phenomena of diffraction.

\section*{Acknowledgments}This work was supported by the National Natural Science Foundation of China (Grants No. 12275212, No. 11447025, and No. 61705182), Shaanxi Fundamental Science Research Project for Mathematics and Physics (Grant No. 22JSY008), the Youth Innovation Team of Shaanxi Universities (Grant No. 24JP177), the Technology Innovation Guidance Special Fund of Shaanxi Province (Grant No. 2024QY-SZX-17), and the Natural science foundation
of Shaanxi Province (Grant No. 2025JC-YBQN-055).

\section*{Data availability} No data were created or analyzed in this study.

\begin{appendix}

\section{Proof of the complete basis $\{|\psi_n(\theta)\rangle\}$} \label{sec:A}
\setcounter{equation}{0}
\renewcommand{\theequation}{A\arabic{equation}}

By using $\langle 1_x | 1_{x'}\rangle = \delta(x-x')$ and the property of Fourier series on $[0,b]$, one has
\begin{equation} \label{eqa-1}
 \langle \psi_m(\theta) |\psi_n(\theta) \rangle = \frac{1}{b} \int_0^b e^{i(k_n - k_m)x} dx = \delta_{mn},
\end{equation}
hence $\{|\psi_n(\theta) \rangle \}$ forms an orthonormal basis.

To show the completeness of the basis, we resort to the inverse Fourier transform of Eq. \eqref{eq8}, that is,
\begin{equation} \label{eqa-2}
 |1_x\rangle = \frac{1}{\sqrt{b}} \sum_n e^{i\phi(x,\theta)} e^{-ik_n x} |\psi_n(\theta)\rangle,
\end{equation}
and combining this with the completeness of $\{|1_x\rangle\}_{x\in[0,b]}$ (see Sec. \ref{sec:2}), we obtain
\begin{equation} \label{eqa-3}
\begin{aligned}
 \hat{I} &= \frac{1}{b} \int_0^b \sum_{m,n} e^{i(k_m - k_n)x} |\psi_n(\theta)\rangle \langle\psi_m(\theta)| dx \\
         &= \frac{1}{b} \sum_{n} \int_0^b |\psi_n(\theta)\rangle \langle\psi_n(\theta)| dx\\
         &= \sum_n |\psi_n(\theta)\rangle \langle \psi_n(\theta)|,
\end{aligned}
\end{equation}
which is precisely the completeness relation expected for $\{|\psi_n(\theta) \rangle \}$.

\section{Proof of Eqs. \eqref{eq9} and \eqref{eq10}} \label{sec:B}
\setcounter{equation}{0}
\renewcommand{\theequation}{B\arabic{equation}}

From Eqs. \eqref{eq4} and \eqref{eq8} one can obtain
\begin{equation} \label{eqb-1}
\begin{aligned}
 \hat{\mathcal{E}}^{(+)} |\psi_n(\theta)\rangle &= \frac{1}{b} \int_0^b \int_0^b e^{i \phi(x,x',\theta)} e^{ik_{n}x} \hat{a}(x) |1_{x'}\rangle dx dx' \\
         &= \frac{1}{b} \int_0^b \int_0^b e^{i \phi(x,x',\theta)} e^{ik_{n}x} \delta_{x,x'} |0\rangle dx dx' \\
         &= \frac{1}{b} \int_0^b e^{ik_{n}x} |0\rangle dx,
\end{aligned}
\end{equation}
where we have defined $\phi(x,x',\theta)=\phi(x,\theta)-\phi(x',\theta)$ and $\delta_{x,x'} =\delta(x-x')$ for notational convenience, and the second equality is due to $\hat{a}(x) |1_{x'}\rangle=\delta(x-x')|0\rangle$.

Then for $n=0$, as $k_{n}=0$, we proved Eq. \eqref{eq9}. For any $n\neq 0$, one always has $\int_0^b e^{ik_{n}x} dx =0$, and this completes the proof of Eq. \eqref{eq10}.

\section{Derivation of $c_n$ in Eq. \eqref{eq14}} \label{sec:C}
\setcounter{equation}{0}
\renewcommand{\theequation}{C\arabic{equation}}

From Eqs. \eqref{eq6} and \eqref{eq14} one can obtain $c_n=\langle \psi_n(\theta)|S\rangle$ ($\forall n$). By combining this with Eq. \eqref{eq8}, we obtain
\begin{equation}\label{eqc-1}
\begin{aligned}
  c_n & = \frac{1}{b} \int_0^b \int_0^b e^{i\phi(x,\theta)e^{-i k_n x}} \langle 1_x|1_{x'}\rangle dx dx' \\
      & = \frac{1}{b} \int_0^b e^{i\phi(x,\theta)}e^{-i k_n x} dx, \\
      & = e^{i(\beta-n\pi)}\frac{\sin(\beta-n\pi)}{\beta-n\pi},
\end{aligned}
\end{equation}
where the second equality is due to $\langle 1_x|1_{x'}\rangle= \delta(x-x')$, and we have denoted by $\beta = (\pi b \sin\theta)/\lambda$ for notational convenience.

\section{Proof of $[\hat{J}_m, \hat{J}_n^\dagger] = \delta_{mn}$} \label{sec:D}
\setcounter{equation}{0}
\renewcommand{\theequation}{D\arabic{equation}}


From the definition of $J_n^\dagger$ in Eq. \eqref{Jn} and the canonical commutation relation $[\hat{a}(x),\hat{a}^\dagger(x')]=\delta(x-x')$, one can obtain that
\begin{equation} \label{eqd-1}
\begin{aligned}
 [\hat{J}_m,\hat{J}_n^\dagger] &= \frac{1}{b} \int_0^b \int_0^b e^{i\phi(x,x',\theta)} e^{i(k_n x'-k_m x)} \delta_{x,x'} dx dx' \\
	                           &= \frac{1}{b} \int_0^b e^{i(k_n-k_m)x}\, dx \\
	                           &= \delta_{mn},
\end{aligned}
\end{equation}
where we have used $\phi(x,x',\theta)$ and $\delta_{x,x'}$ defined below Eq. \eqref{eqb-1} for notational convenience, and the last equality is due to orthogonality of the Fourier modes on the interval $[0,b]$.

Since the bright-state and dark-state operators are given by $\hat{B}_\theta^\dagger= \hat{J}_0^\dagger$ and $\hat{D}_{n,\theta}^\dagger=\hat{J}_n^\dagger$ ($\forall n\neq 0$), respectively, it follows immediately that
\begin{equation} \label{eqd-2}
 [\hat{B}_\theta, \hat{B}_\theta^\dagger] = 1, ~
 [\hat{D}_{m,\theta}, \hat{D}_{n,\theta}^\dagger] = \delta_{mn}, ~
 [\hat{B}_\theta, \hat{D}_{n,\theta}^\dagger] = 0.
\end{equation}
These relations show that the bright- and dark-state operators form a set of independent bosonic modes adapted to the detection direction $\theta$, thereby establishing the self-consistency of the mode decomposition.

\section{Proof of Eqs. \eqref{eq17} and \eqref{eq25}} \label{sec:E}
\setcounter{equation}{0}
\renewcommand{\theequation}{E\arabic{equation}}

From Eq. \eqref{Jn} one can obtain that
\begin{equation} \label{eqe-1}
\hat{a}^\dag(x)= \frac{1}{\sqrt{b}} \sum_n e^{i\phi(x,\theta)e^{-i k_n x}} \hat{J}_n^\dag,
\end{equation}
and it follows that
\begin{equation}\label{eqe-2}
\begin{aligned}
  \hat{A}^\dag  & = \frac{1}{\sqrt{b}} \int_0^b \left(\frac{1}{\sqrt{b}} \sum_n e^{i\phi(x,\theta)}e^{-i k_n x} \hat{J}_n^\dag \right) dx \\
                & = \sum_n \left(\frac{1}{b}\int_0^b e^{i\phi(x,\theta)}e^{-i k_n x} dx \right) \hat{J}_n^\dag, \\
                & = \sum_n c_n \hat{J}_n^\dag,
\end{aligned}
\end{equation}
where the last equality is due to Eq. \eqref{eqc-1}. This, together with $\hat{B}_\theta^\dagger = \hat{J}_0^\dag$ and $\hat{D}_{n,\theta}^\dagger = \hat{J}_n^\dagger$ ($n\neq 0$), completes the proof of Eq. \eqref{eq17}.

Moreover, from Eq. \eqref{eqe-1} one can obtain
\begin{equation} \label{eqe-3}
\begin{aligned}
 \int_0^b \alpha \hat{a}^\dagger(x) dx
  & = \alpha \sum_n \left(\frac{1}{\sqrt{b}}\int_0^b e^{i\phi(x,\theta)e^{-ik_n x}} dx \right) \hat{J}_n^\dag \\
  & = \sum_n \alpha c_n \sqrt{b} \hat{J}_n^\dag,
\end{aligned}
\end{equation}
where the second equality is due to Eq. \eqref{eqc-1}. Similarly, one can obtain
\begin{equation} \label{eqe-4}
 \int_0^b \alpha^{*} \hat{a}(x) dx= \sum_n \alpha^{*} c_n^{*} \sqrt{b} \hat{J}_n.
\end{equation}
This, together with Eq. \eqref{eq24}, completes the proof of Eq. \eqref{eq25}.

\section{Proof of Eq. \eqref{eq26}} \label{sec:F}
\setcounter{equation}{0}
\renewcommand{\theequation}{F\arabic{equation}}

Using the decomposition of the coherent state $|\alpha\rangle$ in Eq. \eqref{eq25}, one immediately obtains the eigenvalue equation for the bright-state operator:
\begin{equation}
\begin{aligned}
 \hat{B}_\theta |\alpha\rangle
  &= \hat{B}_\theta |\alpha_{0}\rangle\otimes_{n\neq 0}|\alpha _{n}\rangle \\
  &= \alpha_0 |\alpha_{0}\rangle\otimes_{n\neq 0}|\alpha_{n}\rangle \\
  &= \alpha_0 |\alpha\rangle,
\end{aligned}
\end{equation}
where the second equality is due to $\hat{B}_\theta = \hat{J}_0$ and $\hat{J}_n |\alpha_n\rangle = \alpha_n |\alpha_n\rangle$. Similarly, for $\hat{D}_{n,\theta}=\hat{J}_n$ ($n\neq 0$) associated with the dark modes, the same decomposition yields
\begin{equation}
\begin{aligned}
 \hat{D}_{n,\theta} |\alpha\rangle
  &= |\alpha _{0}\rangle\otimes \hat{D}_{n,\theta} |\alpha_{n}\rangle\otimes _{m\neq 0,n}|\alpha _{m}\rangle \\
  &= \alpha_n |\alpha_{0}\rangle\otimes |\alpha _{n}\rangle\otimes_{m\neq 0,n}|\alpha_{m}\rangle \\
  &= \alpha_n |\alpha\rangle.
\end{aligned}
\end{equation}

Hence, the multimode coherent state $|\alpha\rangle$ is the eigenstate of the bright-state operator $\hat{B}_\theta$ and of the dark-state operator $\hat{D}_{n,\theta}$ ($\forall n\neq 0$), with eigenvalues $\alpha_0$ and $\alpha_n$, respectively. These relations directly lead to the first-order correlation function in the main text.

\end{appendix}

\end{document}